\documentclass[acmsmall]{acmart}
\usepackage{xcolor}
\usepackage{framed}
\usepackage{changepage}
\usepackage[most]{tcolorbox}
\usepackage{multirow}
\usepackage{xspace}
\usepackage{colortbl}
\usepackage{booktabs}
\usepackage{setspace}
\usepackage{fontawesome5}

\AtBeginDocument{%
  }

\setcopyright{acmlicensed}
\acmDOI{10.1145/3729342}
\acmYear{2025}
\acmJournal{PACMSE}
\acmVolume{2}
\acmNumber{FSE}
\acmArticle{FSE072}
\acmMonth{7}
\received{2025-02-25}
\received[accepted]{2025-04-01}

\begin{document}
\definecolor{PAblue}{RGB}{0,122,204}%
\definecolor{PAlightblue}{RGB}{235, 247, 255}%
\definecolor{mygray}{RGB}{184, 184, 184}%
\definecolor{mygray2}{RGB}{224, 224, 224}%
\newcommand{\icon}[1]{{\includegraphics[height=1.5\fontcharht\font`\B]{#1}}\xspace}
\newcommand{\meiicon}{\icon{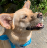}}
\newcommand{\obsGroup}[2]{{\footnotesize{#1}$_#2$}}

\newcommand{\quo}[1]{ 
	\vspace{-0.15cm}
	\def\FrameCommand{%
		\hspace{0pt}%
		{\color{PAblue}\vrule width 2pt}%
		{\color{white}\vrule width 2pt}%
		\colorbox{white}
	}%
	\MakeFramed{\advance\hsize-\width\FrameRestore}%
	\noindent\hspace{-4.55pt}%
	\begin{adjustwidth}{}{0pt}
		\emph{#1}
		\vspace{-3pt}
	\end{adjustwidth}\endMakeFramed%
}

\newcommand{\personquo}[2]{ 
	\vspace{-0.15cm}
	\def\FrameCommand{%
		\hspace{0pt}%
		{\color{PAblue}\vrule width 2pt}%
		{\color{white}\vrule width 2pt}%
		\colorbox{white}
	}%
	\MakeFramed{\advance\hsize-\width\FrameRestore}%
	\noindent\hspace{-4.55pt}%
	\begin{adjustwidth}{}{0pt}
		``\emph{#1}'' \newline ---{#2}
		\vspace{-3pt}
	\end{adjustwidth}\endMakeFramed%
}

\definecolor{boxcolor}{RGB}{238, 223, 204} %
\DeclareRobustCommand{\mybox}[2][gray!20]{%
\begin{tcolorbox}[   %
        breakable,
        left=0pt,
        right=0pt,
        top=0pt,
        bottom=0pt,
        colback=#1,
        colframe=black,
        width=\dimexpr\columnwidth\relax, 
        enlarge left by=0mm,
        boxsep=5pt,
        outer arc=4pt,
        boxrule=.5mm
        ]
        #2
\end{tcolorbox}
}

\newcommand{\pquote}[2]{{\emph{``#1''} (P#2)}}

\newcommand{\todo}[1]{{\color{orange} \bfseries TODO: #1}}
\newcommand{\added}[1]{{\color{black} #1}}

\newcommand{\theme}[1]{\emph{#1}}
\newtcbox{\chip}[1][]{enhanced,
 box align=base,
 nobeforeafter,
 colback=mygray2,
 colframe=mygray2,
 fontupper=\scriptsize\ttfamily,
 left=0.2pt,
 right=0.2pt,
 top=0.2pt,
 bottom=0.2pt,
 boxsep=0.4pt,
 #1}

\title{Prompts Are Programs Too! Understanding How Developers Build Software Containing Prompts}

\author{Jenny T. Liang}
\orcid{0000-0001-6722-9959}
\affiliation{%
  \institution{Carnegie Mellon University}
  \city{Pittsburgh}
  \country{USA}
}
\email{jtliang@cs.cmu.edu}

\author{Melissa Lin}
\authornote{Authors contributed equally to this research.}
\orcid{0009-0001-6795-8040}
\affiliation{%
  \institution{Carnegie Mellon University}
  \city{Pittsburgh}
  \country{USA}
}
\email{mylin@andrew.cmu.edu}

\author{Nikitha Rao}
\authornotemark[1]
\orcid{0009-0004-5404-1534}
\affiliation{%
  \institution{Carnegie Mellon University}
  \city{Pittsburgh}
  \country{USA}
}
\email{nikitharao@cmu.edu}

\author{Brad Myers}
\orcid{0000-0002-4769-0219}
\affiliation{%
  \institution{Carnegie Mellon University}
  \city{Pittsburgh}
  \country{USA}
}
\email{bam@cs.cmu.edu}

\renewcommand{\shortauthors}{Liang, Lin, Rao, Myers}

\begin{abstract}
Generative pre-trained models power intelligent software features used by millions of users controlled by developer-written natural language prompts.
Despite the impact of prompt-powered software, little is known about its development process and its relationship to programming.
In this work, we argue that some prompts are programs and that the development of prompts is a distinct phenomenon in programming known as \emph{``prompt programming''}.
We develop an understanding of prompt programming using Straussian grounded theory through interviews with 20 developers engaged in prompt development across a variety of contexts, models, domains, and prompt structures.
We contribute 15 observations to form a preliminary understanding of current prompt programming practices.
For example, rather than building mental models of code, prompt programmers develop mental models of the foundation model (FM)'s behavior on the prompt by interacting with the FM.
While prior research shows that experts have well-formed mental models, we find that prompt programmers who have developed dozens of prompts still struggle to develop reliable mental models.
Our observations show that prompt programming differs from traditional software development, motivating the creation of prompt programming tools and providing implications for software engineering stakeholders.

\end{abstract}

\ccsdesc[500]{Software and its engineering~Software development methods}
\ccsdesc[100]{Human-centered computing~Empirical studies in HCI}
\ccsdesc[300]{Computing methodologies~Artificial intelligence}

\keywords{Prompt programming, prompt engineering, Straussian grounded theory}

\maketitle

\section{Introduction}
\label{sec:introduction}
\personquo{I suspect that machines to be programmed in our native tongues---be it Dutch, English, American, French, German, or Swahili---are as damned difficult to make as they would be to use.}{Edsger W. Dijkstra (1979)}

Generative pre-trained models (e.g., GPT-4~\cite{achiam2023gpt}, Dall-E~\cite{betker2023improving})---also known as foundation models (FMs)---have changed how programmers build software.
AI programming assistants that generate code (e.g., GitHub Copilot~\cite{chen2021evaluating}) have improved developer productivity~\cite{ziegler2022productivity,peng2023impact, ziegler2024measuring} by helping developers write significant portions of code, learn new APIs and programming languages, and write tests~\cite{liang2024large}.
Recently, instruction-tuned FMs~\cite{ouyang2022training} like ChatGPT~\cite{openai2024chatgpt} have expanded the assistance with software development tasks by writing natural language prompts.
This includes resolving code issues, developing new features,  refactoring, and information seeking~\cite{chavan2024analyzing, haque2024information}.

The rising prominence of prompts has ushered in \textit{``prompt engineering''}, whereby FM users repeatedly write and revise natural language prompts.
These prompts enable new intelligent features when integrated in popular software applications~\cite{parnin2023building}, such as Google Search~\cite{google2024google} and Microsoft Office~\cite{microsoft2024copilot}, reaching millions of users~\cite{microsoft2024bringing}.
As of January 2024, engineered prompts have also powered over 3 million custom versions of ChatGPT for specific tasks, known as GPTs~\cite{openai2024introducing}.

Yet, little is known about prompt engineering and its relationship to programming.
Relevant works include \citet{dolata2024development}'s study of 52 freelance developers building solutions based on generative AI and \citet{parnin2023building}'s study of 26 professional software developers integrating generative AI into products.
While these studies offer insight on prompt development, they are constrained to freelance and professional software development, limiting the generalizability of the results to other programming contexts.
We address these gaps by following a more systematic and rigorous qualitative methodology---Straussian grounded theory~\cite{corbin2015basics}---to study a diverse sample of programmers to understand the process of developing prompts embedded in software applications.

In this work, we argue that some prompts function as programs, a phenomenon we call \emph{prompt programming}~\cite{jiang2022promptmaker}.
In prompt programming, prompts are authored at design-time and executed at runtime with variable user inputs, but are written in natural language instead of a programming language.
Prompt programming is a specific form of prompt engineering---broadly defined as iterating on prompts to improve FM outputs~\cite{zamfirescu2023johnny}---where the same prompt handles variable inputs given at runtime. 
For example, a prompt program on a teleconferencing platform may take a meeting transcript and attendee name and return bulleted todos and deadlines for that individual.

In prompt programming, natural language is a programming abstraction, but unlike formalisms such as DSLs or machine code, it introduces ambiguity and reduces precision~\cite{dijkstra2005foolishness}.
While the idea of programming in natural language dates as early as the 1970s, where Dijkstra discussed a vision of instructing machines in ``our native tongues''~\cite{dijkstra2005foolishness}, it has become feasible and widespread with FMs due to instruction-tuning~\cite{ouyang2022training} that teaches models to follow natural language instructions.

We argue that prompt programming is a phenomenon that warrants its own study, as the sociotechnical circumstance of the programming task influences the nature of the programming.
For example, the programming process of end-users who write software for personal use (i.e., end-user programming~\cite{ko2011state}) is different from that of data scientists who write software to explore different possibilities in code (i.e., exploratory programming~\cite{kery2017exploring, sheil1986datamation}), and both differ from a professional programmer who focuses on a small component within a larger system.
The programming task also varies with the developer's tools (e.g., visual programming languages vs. Jupyter notebooks), focus on the formality of the process (e.g., accomplishing a goal vs. exploring ideas), and key challenges (e.g., finding the right abstractions vs. managing code versions)~\cite{ko2011state, kery2017exploring}. 
This creates disparate experiences and necessitates the development of support tools unique to each context.
Thus, studying prompt programs could inform the development of tools to support prompt programmers.

\begin{figure}
    \begin{tcolorbox}[colback=PAlightblue,colframe=PAblue,title=\textbf{\hspace{1.5em} Study Findings: Barriers in Prompt Programming},left=-3pt,right=8pt,top=4pt,bottom=4pt,fontupper=\small, fonttitle=\small]
      \begin{itemize}
        \item[] \textbf{Programmer} (Section~\ref{sec:programmer})
        \item[1.] \obsGroup{\faLightbulb}{{}} Programmers must develop a mental model about the FM's behavior on the prompt. 
        \item[2.] \obsGroup{\faLightbulb}{{}} Programmers' mental models are not reliable. 
        \item[3.] \obsGroup{\faLightbulb}{{}} Programmers use external knowledge sources and prior experience to build their mental model. \\

        \item[] \textbf{Foundation Model} (Section~\ref{sec:model})
        \item[4.] \obsGroup{\faLightbulb}{{}} Each FM has its own set of qualities and capabilities.\\

        \item[] \textbf{Prompt} (Section~\ref{sec:prompt})
        \item[5.] \obsGroup{\faDiceSix}{{}} Minute details in the prompt matter. 
        \item[6.] \obsGroup{\faDiceSix}{{}} Prompts are finicky and fragile. \\

        \item[] \textbf{Requirements} (Section~\ref{sec:requirements})
        \item[7.] \obsGroup{\faFileCode}{{}} Assumptions of the requirements must be explicitly stated.
        \item[8.] \obsGroup{\faLightbulb}{{}} Requirements can evolve as the capabilities of the FM are discovered. \\

        \item[] \textbf{Design} (Section~\ref{sec:design})
        \item[9.] \obsGroup{\faFileCode}{{}} Developers make decisions about how prompt programs should be composed and decomposed.
        \item[10.] \obsGroup{\faLightbulb}{{}} Dependent prompts are tightly coupled. \\

        \item[] \textbf{Implementation} (Section~\ref{sec:implementation})
        \item[11.] \obsGroup{\faLightbulb}{{}} Prompt programming is rapid and unsystematic. \\

        \item[] \textbf{Debugging} (Section~\ref{sec:debugging})
        \item[12.] \obsGroup{\faDiceSix}{{}} Fault localization is difficult. \\

        \item[] \textbf{Data Curation} (Section~\ref{sec:data-curation})
        \item[13.] \obsGroup{\faClipboardCheck}{{}} Programmers need to find representative data for the task.\\

        \item[] \textbf{Evaluation} (Section~\ref{sec:evaluation})
        \item[14.] \obsGroup{\faClipboardCheck}{{}} Evaluating prompt programs requires assessing qualitative constructs.
        \item[15.] \obsGroup{\faClipboardCheck}{{}} Testing occurs at different scopes.
    \end{itemize}
    \end{tcolorbox}
    \vspace{-0.5\baselineskip}
    \caption{
    An overview of the 15 study findings on the barriers in prompt programming.
    These can be divided into four types of barriers: understanding FM behavior (\obsGroup{\faLightbulb}{{}}), dealing with stochasticity (\obsGroup{\faDiceSix}{{}}), programming in natural language (\obsGroup{\faFileCode}{{}}), and testing prompt program behavior (\obsGroup{\faClipboardCheck}{{}}).
    }
    \label{fig:findings}
\end{figure}

We form a preliminary understanding of current prompt programming practices using Straussian grounded theory~\cite{corbin2015basics}, a qualitative research methodology that develops a novel exploratory explanation of a domain known as a ``grounded theory''.
To this end, we interview 20 developers building software that is based on prompts to understand: \emph{How do programmers develop programs that incorporate natural language prompts?}
First, we consider the definitions of a \emph{prompt} and \emph{program}.
We define a \emph{prompt} as:
\vspace{-1.75em}
\quo{A natural language query to a FM.}

For the definition of a \emph{program}, we consider \citet{ko2011state}'s definition, which is ``a collection of specifications that may take variable inputs, and that can be executed (or interpreted) by a device with computational capabilities.''
We extend this to derive the definition of a \textit{prompt program}:
\quo{A prompt that accepts variable inputs and could be interpreted by a FM to perform specified actions and/or generate output. This prompt is executed within a software application or code by a FM.}

This definition includes developing prompts for applications like GPTs and chatbots.  
It also covers language agents, where large language models (LLM) receive arbitrary user requests and execute them in dynamic environments using external modules known as ``tools" (e.g., modules for mathematical reasoning~\cite{schick2024toolformer} and code interpreters~\cite{gao2023pal}). 
However, this definition excludes cases where a developer converses with an FM once to achieve a task (e.g., prompt engineering with ChatGPT to debug a specific error message). 
These prompts do not accept variable inputs, do not generalize beyond the specific case, and are not executed within a broader software application.

Through our investigation, we contribute 15 observations that identify barriers in prompt programming (see Figure~\ref{fig:findings}).
These form a preliminary understanding of current prompt programming practices.
To aid discussion, we organize these into four types of barriers: understanding FM behavior (\obsGroup{\faLightbulb}{{}}), dealing with stochasticity (\obsGroup{\faDiceSix}{{}}), programming in natural language (\obsGroup{\faFileCode}{{}}), and testing prompt program behavior (\obsGroup{\faClipboardCheck}{{}}).
While some results corroborate those from \citet{dolata2024development}'s and parnin2023building's studies (e.g., fault localization being difficult (\obsGroup{\faDiceSix}{{\#12}})), our results differ in notable ways.
First, we find prompt programmers interact with FMs to develop mental models of the FM's behavior on the prompt (\obsGroup{\faLightbulb}{{\#1}}) and its unique qualities (\obsGroup{\faLightbulb}{{\#4}}).
Although the literature suggests that experts have well-formed mental models compared to novices~\cite{staggers1993mental},
prompt programmers struggle to develop reliable mental models (\obsGroup{\faLightbulb}{{\#2}}) even after writing dozens of prompts, each with many iterations.
This contributes to a rapid and unsystematic development process (\obsGroup{\faLightbulb}{{\#11}}).
We also find that prompts can be composed and decomposed (\obsGroup{\faFileCode}{{\#9}}) and testing occurs at different scopes (\obsGroup{\faClipboardCheck}{{\#15}}). 
While these observations exhibit similarities to software development, we also find differences compared to the process of developing traditional software, necessitating the creation of new tools and processes to support prompt programming.
This has implications for software practitioners, academics, tool creators, and educators interested in generative AI in software engineering.

\section{Related Work}
We discuss related work on prompt engineering (Section~\ref{sec:prompt-engineering}) and software engineering for AI (SE4AI; see Section~\ref{sec:human-aspects-se4ai}).
Due to the rapidly evolving landscape of prompt engineering and FMs, this discussion will necessarily miss more recent developments.

\subsection{Prompt Engineering}
\label{sec:prompt-engineering}

Prior literature has examined prompt engineering in a variety of contexts.
This includes user studies on prompt engineering, prompt engineering tools, and retrospective interview studies on prompt engineering in software development.
We discuss this in more detail below.

Previous work has conducted empirical user studies of prompt engineering.
Some works have focused on the development of chatbots with end users using prompts~\cite{zamfirescu2023conversation, zamfirescu2023herding, zamfirescu2023johnny}.
In a user study of 10 people who used a tool to create chatbots through prompts, \citet{zamfirescu2023johnny} identified several challenges of prompt development, including opportunistic prompt design approaches, a lack of systematic testing, and writing prompts that were not generalized.
\citet{jiang2022promptmaker} evaluated a prompt development tool in 11 users in various roles, including designers, content strategists, and front-end developers. 
The problems participants faced included the prompt easily breaking and overfitting on examples, as well as having difficulty evaluating large amounts of text.

Other work has investigated the creation of prompting tools to assist with various prompting tasks.
Prompt pattern catalogs can solve common problems encountered when conversing with an LLM~\cite{white2023prompt}, such as question refinement, as well as software engineering tasks~\cite{white2024chatgpt}, such as specification disambiguation.
Other relevant tools include prompting interfaces, such as PromptAid~\cite{mishra2023promptaid}, PromptMaker~\cite{jiang2022promptmaker}, and PromptIDE~\cite{strobelt2022interactive}.
These tools included separate views to show the dataset, iterate on the prompt, track the performance of the prompt, and search for the prompts.

Most related to this study are retrospective interview studies that investigate prompt development in software engineering.
\citet{dolata2024development} conducted a study with 52 freelance developers on their experience with building prompt-powered software.
Their interview focuses on the positive and negative experiences with generative AI, project uncertainties, and views on freelancing.
The methodology involved recruiting from Upwork, conducting retrospective interviews, performing a thematic analysis, and comparing the findings with four meta-reviews of SE4AI literature.
The participants enumerated several challenges, including having difficulty identifying the source of incorrect responses, budget constraints, and unrealistic client expectations.
Meanwhile, \citet{parnin2023building} conducted an interview study with 26 professional programmers who developed product copilots.
Interview topics included motivation for using AI in products, major tasks for building the generative AI application, prompt engineering, testing, tooling, challenges, learning-related skills, and concerns with AI.
Their methodology involved recruiting participants from UserInterviews.com, conducting retrospective interviews and structured brainstorming sessions, and performing a thematic analysis.
They found that the participants followed a general process of exploration, implementation, evaluation, and productization.
This work noted several challenges, including the trial-and-error nature of prompt development and creating benchmarks.

This literature provides an initial understanding of the unique aspects of prompt programming.
However, they are constrained to specific types of developers (i.e., end-users, professional programmers, and freelancers), limiting the generalizability of the results.
Additionally, these works do not study the process of writing prompt programs with respect to software development activities.
Our study builds upon these works by employing a more systematic and rigorous qualitative methodology, Straussian grounded theory~\cite{corbin2015basics}, to study the process of writing prompt programs.
We sample a broad range of prompt programmers along various axes, such as prompt structure, programming context, and role, as well as ask questions about the end-to-end development process.

\subsection{Software Engineering for AI}
\label{sec:human-aspects-se4ai}
Numerous works have studied the human aspects of building ML-enabled systems, also known as software engineering for AI.
These works have documented the experiences of practitioners and their common challenges in this domain.
We elaborate further below.

Building ML-enabled systems involves working with and wrangling nondeterministic, opaque neural models~\cite{giray2021software}. 
This is similar to prompt programming, as developers must also contend with non-transparent, stochastic FMs.
Previous work describes the development of models as highly experimental~\cite{giray2021software, nahar2022collaboration, nahar2023meta, wan2019does, wolf2020sensemaking}.
The literature also stresses the importance of collecting high-quality data for the domain~\cite{nahar2022collaboration, giray2021software, dolata2024development, arpteg2018software, qian2024understanding, nahar2023meta} as well as the use of quantitative metrics, such as precision, precision, and recall, to measure model performance~\cite{wan2019does, nahar2022collaboration}.

However, the development of ML components presents unique challenges.
Models can be difficult to debug due to their determinism~\cite{giray2021software}, while \emph{training-serving skew} can arise when training data fail to generalize to production~\cite{nahar2022collaboration, nahar2023meta, wan2019does}. 
To mitigate this, models are periodically re-trained~\cite{nahar2023meta}.
In addition, expertise in a variety of domains, such as software engineering and data science, is distributed between roles~\cite{nahar2022collaboration, wan2019does, giray2021software}.
In an interview study with 45 practitioners, \citet{nahar2022collaboration} found this introduced collaboration challenges between different roles, such as unclear model requirements, handling evolving data, and inadequate datasets.
Finally, a lack of AI literacy can make requirements elicitation, communication, and collaboration with clients challenging~\cite{nahar2022collaboration, dolata2023making, wolf2020sensemaking}.

This body of literature serves as a foundation for understanding prompt programming.
We extend this body of work by comparing the building of ML-enabled systems to prompt programming and understanding what aspects of this prior literature apply to prompt programming.

\begin{figure}[t!]
\centering
\includegraphics[trim=0 550 0 0, clip, width=0.92\linewidth, keepaspectratio]{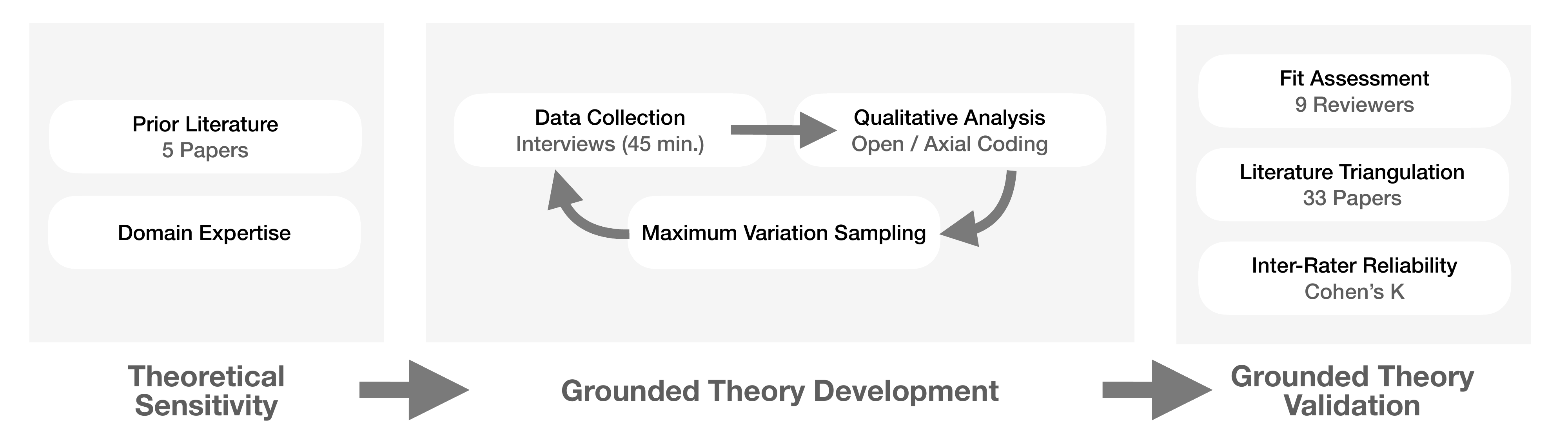} 
\caption{
An overview of the Straussian grounded theory methodology performed in the study.
}
\label{fig:methodology}
\end{figure}

\section{Methodology}
Since there is limited literature on prompt programming, we used a qualitative approach to explore the \emph{process} of developing prompt-powered software.
We used the grounded theory methodology, which allows researchers to develop a novel exploratory explanation for a domain.

There are three popular approaches to grounded theory: Glaserian/classical~\cite{glaser2017discovery}, Straussian~\cite{corbin2015basics}, and constructivist~\cite{charmaz2006constructing} methodologies, which vary in their procedures and epistemology~\cite{cole2022more}. 
Our approach is based on Straussian grounded theory~\cite{corbin2015basics} (see Figure~\ref{fig:methodology}). 
This involves defining a research question; developing theoretical sensitivity to the phenomenon through prior literature and domain expertise; generating a grounded theory via diverse sampling and simultaneous data collection and analysis; and validating the grounded theory by triangulating with literature, performing an assessment of the fit of the grounded theory, and computing inter-rater reliability~\cite{cole2022more, corbin2015basics}.
In this section, we first discuss our grounded theory process (Section~\ref{sec:grounded-theory-process}).
We then describe our participants (Section~\ref{sec:participants}) and interview protocol (Section~\ref{sec:protocol}), and close with a discussion of the limitations of our method (Section~\ref{sec:threats-to-validity}). 
We present our results in Section~\ref{sec:theory} and Section~\ref{sec:process}.

\subsection{Grounded Theory Process}
\label{sec:grounded-theory-process}

We describe our grounded theory process below.
It contains five main stages: defining a research question, developing theoretical sensitivity, generating the grounded theory, triangulating with the literature, and validating the grounded theory.
We describe this process in further detail below.

\paragraph{Defining a research question}
Our process began with defining a research question.
We study the phenomenon where developers write a prompt program using natural language prompts, rather than pure code: \emph{How do programmers develop programs that incorporate natural language prompts?}
Prompts can be used in a variety of contexts, such as interacting with ChatGPT in conversation~\cite{chavan2024analyzing}, but are not programs.
Thus, we apply the definition of a prompt program described in Section~\ref{sec:introduction}.

\paragraph{Developing theoretical sensitivity}
Before the study, the investigator should develop an intuition for the phenomenon being studied, known as \emph{theoretical sensitivity}~\cite{cole2022more}.
This can be done through professional experience and understanding the prior literature so that researchers can extract insights from the data.
However, the investigators should not be constrained by prior knowledge while developing the grounded theory~\cite{corbin2015basics}.
We entered the study with some theoretical sensitivity since two authors are software engineering and AI researchers and have developed multiple prompt programs.
Following prior work~\cite{mcghee2007grounded}, we considered our background and opted for a lightweight overview of prompting literature.
We considered foundational prompting papers in machine learning venues~\cite{brown2020language, wei2022chain}, a notable survey paper on prompting techniques in natural language processing (NLP)~\cite{liu2023pre} as well as two empirical studies on how people develop prompts~\cite{zamfirescu2023johnny, dolata2024development} from human-computer interaction (HCI) and software engineering venues.
We considered this literature and the authors' existing domain expertise to develop the interview protocol (Section~\ref{sec:protocol}).

\paragraph{Generating the grounded theory}
We conducted 45-minute interviews with a diverse set of 20 prompt programmers (see Table~\ref{tab:participants}) over Zoom.
The interviews were recorded and transcribed; the recordings were later deleted.
Two authors were present for the first six interviews to become familiar with the data.
The first author conducted the remaining interviews.

To develop the initial grounded theory, three authors first independently performed line-by-line coding on the first two interviews in separate codebooks to become more sensitized to the data.
Each code contained a description of the code as well as observations from the interviews.
The authors then reconvened to merge the individual codebooks by identifying codes with similar concepts and merging them into a shared codebook.
The remaining codes were then discussed and added or removed to the codebook by unanimous vote.
The first author then coded the next interview, and the other authors reviewed the codes for agreement.
We identified five instances of disagreement, which were discussed and resolved.
The three authors then performed axial coding, grouping the emerging codes into preliminary categories on unanimous vote.

For the remaining interviews, the authors individually open coded the interview transcripts, noting new codes that emerged in the shared codebook.
A memo was created for each interview to capture key insights and refine the emerging grounded theory.
The authors met regularly to discuss the new codes and observations.
New codes and observations were added to the shared codebook upon unanimous vote, and the categories were further refined as more data was collected.

As the grounded theory developed, we performed maximum variation sampling~\cite{suri2011purposeful} to obtain a diverse set of participants that could challenge or extend the grounded theory.
We recruited participants who met the definition of creating a prompt program (Section~\ref{sec:introduction}) by snowball sampling within the authors' social networks and recruiting in online open-source communities.
Participants were recruited based on the types of models used (e.g., open-source models vs. closed-source models, vision language models vs. language models), application domain (e.g., robotics, education, productivity), organization size (e.g., large technology company vs. startup), programming context (e.g., academia, industry, freelance, open-source software), role, and/or prompt structure (e.g., a language agent that is able to use tools vs. single prompt vs. multi-prompt).
We initially achieved theoretical saturation after interviewing 14 participants.
We continued to recruit participants to ensure that our observations held across a variety of situations, including open-source software and startups.
We stopped data collection after reaching 20 participants.

\begin{table*}
  \centering \footnotesize
\caption{
An overview of the participants in the study. 
We report the participant's development context, number of years of programming experience, number of prompt programs developed prior to the interview, prompt structure, model type, application domain, and task.
Prompt structures can be a single prompt, multiple chained prompts (multi-prompt), or a language agent (agent).
Model types can be open source (\chip{OS}) or closed source (\chip{CS}) and can be a language model (\chip{LM}) or a vision-language model (\chip{VLM}).
}

\label{tab:participants}
\begin{tabular}{p{0.02\linewidth}p{0.12\linewidth}p{0.06\linewidth}p{0.09\linewidth}p{0.12\linewidth}p{0.07\linewidth}p{0.145\linewidth}p{0.175\linewidth}}
\toprule
 \textbf{ID} & \textbf{Context} & \textbf{Exp.} & \textbf{\# Prompts} & \textbf{\added{Structure}} & \textbf{Model} & \textbf{App Domain} & \textbf{Task} \\
\midrule
P1 & Academia & 16 yrs & 30 &  Single prompt & \chip{OS} \chip{LM} & Security & Classification \\
\addlinespace[1pt]
{P2} & \parbox[t][2em][t]{0.12\linewidth}{\setstretch{0.8} Personal project} & {10 yrs} & {40} & {Single prompt} & {\chip{CS} \chip{LM}} & {Productivity} & \parbox[t][2em][t]{2.5cm}{\setstretch{0.8} Text summarization (chatbot)} \\
\addlinespace[1pt]
P3 & Academia & 8 yrs & 30 &  Single prompt & \chip{CS} \chip{LM} & Research analysis & Classification \\
\addlinespace[1pt]
P4 & Big tech (R\&D) & 8 yrs & 10 &  Single prompt & \chip{OS} \chip{LM} & Software testing & Code generation \\
\addlinespace[1pt]
P5 & Academia & 14 yrs & 35 &  Single prompt & \chip{CS} \chip{LM} & Education & Generation \\
\addlinespace[1pt]
P6 & Academia & 10 yrs & 3 &  Multi-prompt & \chip{CS} \chip{LM} & Software testing & Code generation \\
\addlinespace[1pt]
P7 & Big tech (Eng) & 10 yrs & 4 &  Single prompt & \chip{CS} \chip{LM} & Education & Generation \\
\addlinespace[1pt]
P8 & Academia & 10 yrs & 20 &  Agent & \chip{CS} \chip{LM}& Software testing & Code generation \\
\addlinespace[1pt]
P9 & Academia & 14 yrs & 20 &  Multi-prompt & \chip{OS} \chip{VLM} & Visual Q\&A & Retrieval, Q\&A \\
\addlinespace[1pt] 
P10 & Big tech (Eng) & 7 yrs & 10 &  Multi-prompt & \chip{CS} \chip{LM} & Education & Text editing \\
\addlinespace[1pt]
P11 & Big tech (R\&D) & 8 yrs & 5 &  Agent & \chip{CS} \chip{LM} & Productivity & Data generation \\
\addlinespace[1pt]
P12 & Big tech (Eng) & 14 yrs & 100+ &  Multi-prompt & \chip{CS} \chip{LM} & Productivity & Code summarization \\
\addlinespace[1pt]
P13 & Big tech (Eng) & 20 yrs & 6 &  Multi-prompt & \chip{CS} \chip{LM} & Productivity & Code generation \\
\addlinespace[1pt]
P14 & Big tech (Eng) & 5 yrs & 3 &  Single prompt & \chip{CS} \chip{LM} & Productivity & Code generation \\
\addlinespace[1pt]
P15 & Big tech (Eng) &  5 yrs & 3 &  Single prompt & \chip{CS} \chip{LM} & Career & Q\&A (chatbot) \\
\addlinespace[1pt]
P16 & OSS & 8 yrs & --- &  Agent & \chip{CS} \chip{LM} & Software dev. & End-to-end workflow \\
\addlinespace[1pt]
P17 & Startup (R\&D) & 6 yrs & 5 &  Single prompt & \chip{CS} \chip{LM} & Robotics & Code generation \\
\addlinespace[1pt]
{P18} &  \parbox[t][2em][t]{1.6cm}{\setstretch{0.8} Startup (Eng), OSS} & {3.5 yrs} & {15} & {Agent} & {\chip{CS} \chip{LM}} & {Software testing} & {Test generation} \\
 \addlinespace[1pt]
P19 & Startup (Eng) & 3 yrs & 5 &  Agent & \chip{CS} \chip{LM} & Research analysis & End-to-end workflow \\
\addlinespace[1pt]
P20 & Freelance & 9 yrs & 1 &  Single prompt & \chip{CS} \chip{LM} & Literature & Q\&A (chatbot) \\
\bottomrule
\end{tabular}
\end{table*}

\paragraph{Triangulating with literature}
While developing the grounded theory of prompt programming, we triangulated our findings with the prior literature.
Since the relevant literature was not identifiable through keyword search, we followed previous work~\cite{nahar2022collaboration} in performing a focused, best-effort literature review by performing forward and/or backward snowballing on the initial set of five papers.
The relevant papers were first identified based on titles related to building AI systems or prompt engineering.
They were later verified by reading the manuscript.
In the end, this literature spanned several communities, including NLP, HCI, programming languages, software engineering, and machine learning.
Of the 502 papers considered, we identified 33 as relevant.
See the supplemental materials for this set of papers~\cite{supplemental-materials}.
For each paper, the first author identified text related to prompt programming and applied codes from the codebook, updating the codes as necessary to further refine the grounded theory.

\paragraph{Validating the grounded theory}
To validate the grounded theory, we assess its \emph{fit}.
Straussian grounded theory suggests validating the grounded theory with professionals and study participants to see if they resonate with the findings~\cite{corbin2015basics}.
To this end, we sent an email containing a summary of the grounded theory as well as a draft of Section~\ref{sec:theory}, Section~\ref{sec:process}, and Table~\ref{tab:participants} to the 20 participants and one external researcher who studies prompt programming.
We asked them to reflect on the fit of the grounded theory and provide feedback based on their experience.
Nine people responded to this inquiry and indicated general agreement with the findings, some with wording changes.

Additionally, to assess the reliability of the coding, the first author randomly selected and open-coded two interviews at the category level.
The labels were then removed, with the original highlighted text spans remaining intact.
One author was assigned to each interview and applied the code categories to each highlighted text span.
Based on this procedure, our inter-rater reliability was 0.89 and 0.81 based on Cohen's $\kappa$---indicating that both are almost in perfect agreement~\cite{landis1977measurement}.

\subsection{Participants}
\label{sec:participants}
A summary of our participants is in Table~\ref{tab:participants}.
The inclusion criterion was developers who had previously created at least one prompt program, according to the definition presented in Section~\ref{sec:introduction}.
Our participants were men ($N=17$) and women ($N=3$) from a wide range of technology roles, including Machine Learning Engineer, Senior Software Engineer, Ph.D. Student, and Principal Data and Applied Scientist.
All participants had prior programming experience (median 8.5 years) and had written a prompt program before (median 10 such programs).
Participants also regularly used foundation models: 10 participants reported using them more than once a day, five participants reported using them once daily, and five participants reported using them weekly.

\begin{figure}
\begin{tcolorbox}[colback=PAlightblue,colframe=PAblue,title=\textbf{\hspace{1.65em} Interview Questions},left=-12pt,right=8pt,top=4pt,bottom=4pt, fontupper=\small, fonttitle=\small]
      \begin{itemize}
        \item Think of the most recent time you wrote such a prompt. Do you have one in mind?
        \item Do you currently have access to the prompt? If so, could you please pull up the prompt and any associated history (e.g., on ChatGPT/Bard UI or GitHub), if applicable?
        \item Did you experience any difficulties while using existing tools to develop a prompt?
        \item Did you refer to external information sources to develop the prompt? If so, which ones?
        \item Did you ever refer to a previous version of the prompt as you were developing a new version?
        \item Did you use any metrics or heuristics to determine when the prompt was successful? 
        \item How did you identify what caused an incorrect output? Were you confident in your answer?
        \item Did your prompt change after deployment?
    \end{itemize}
    \end{tcolorbox}
    \vspace{-0.5\baselineskip}
    \caption{A subset of the interview questions. The full protocol is in the supplemental materials~\cite{supplemental-materials}}.
    \label{fig:interview-questions}
\end{figure}

\subsection{Interview Protocol}
\label{sec:protocol}
Interviews were 45 minutes long and conducted over Zoom.
The procedure was approved by our Institution's Review Board.
Participants were compensated with a \$20 USD Amazon gift certificate.

To begin, the participant completed a demographic and background survey with questions such as gender, the number of years of programming experience, and number of prompt programs developed.
We follow best practices in reporting gender in HCI~\cite{scheuerman2020hci}.
The interviewer then presented the participant with the definition of a prompt program and asked the participant to recall the last time they wrote such a program.
If the participant had access to the prompt, they retrieved the prompt and any associated history, and provided a brief overview of the prompt.
17 of the 20 participants accessed their prompts. 
P15, P17, and P19 did not due to privacy concerns or limited access to workplace resources.
To re-familiarize the participant with their prompt and their process, the participant discussed the prompt's design choices (i.e., a decision made in how to implement the prompt) and discussed the challenges they faced while developing the prompt.
Finally, the participant discussed the overall process that they used to develop the prompt program.

To develop the interview protocol, we identified themes to explore: data, requirements, design, implementation, evaluation, debugging, and deployment, based on the literature and the authors' prior experience developing prompt programs (Section~\ref{sec:grounded-theory-process}).
We generated questions for each theme in the protocol.
The challenges were discussed before the process, so they were not biased by the discussion about the prompting process.
A sample of the interview questions are in Figure~\ref{fig:interview-questions}; the complete protocol and demographic and background survey are in the supplemental materials~\cite{supplemental-materials}.

Following best practices in software engineering research~\cite{ko2015practical}, we piloted the  protocol with three individuals who had developed prompt programs to clarify the wording of the survey and the interview and ensure that our definition of a prompt program was clear.

\subsection{Threats to Validity}
\label{sec:threats-to-validity}
Below, we discuss the threats to validity of this study.

\paragraph{Internal validity}
Memory biases could introduce errors in describing prompt design choices and development experience.
To reduce this threat, participants recounted their most recent prompt that met the definition of a prompt program.
When possible, the participants reviewed their prompt.
Although P15, P17, and P19 did not do so, which could produce different data, this threat had limited impact, as their data did not introduce new codes.
In addition, the authors' experience with prompt programming and the interview structure could introduce confirmation biases to the generated categories and the developed grounded theory.
We reduced these threats by performing triangulation with literature and validating the grounded theory with study participants and an external professional who had engaged in prompt programming research.
However, the literature sampling strategy was not fully systematic, which could introduce selection bias in relevant studies.

\paragraph{External validity}
The results of this study may not generalize beyond our sample.
The participants may not be representative of all prompt programmers as snowball sampling, recruiting within the authors' social networks, and self-selection bias could introduce sampling bias.
Study interviews were conducted in English, which could cause less representation from nonnative speakers.
Some participants were unable to disclose all details of their process due to company policy.
In addition, participants shared recent experiences with prompt programming, which could limit the generalizability of our findings. 
Since FM performance evolves over time~\cite{chen2023chatgpt}, observations related to model behavior (e.g., prompt sensitivity) may not remain consistent.

\paragraph{Conclusion validity}
Qualitative studies rely on the interpretation of the participant interview data, which introduces subjectivity and the potential for bias in the analysis. 
To mitigate this threat, all authors involved in the analysis attended multiple interview sessions to familiarize themselves with the data. 
The grounded theory was iteratively developed through unanimous agreement among researchers. 
Additionally, we assessed inter-rater reliability and conducted validation with eight participants and an external expert to ensure the credibility of our findings.

\begin{figure}[t!]
\centering
\includegraphics[trim=100 190 110 250, clip, width=0.9\linewidth, keepaspectratio]{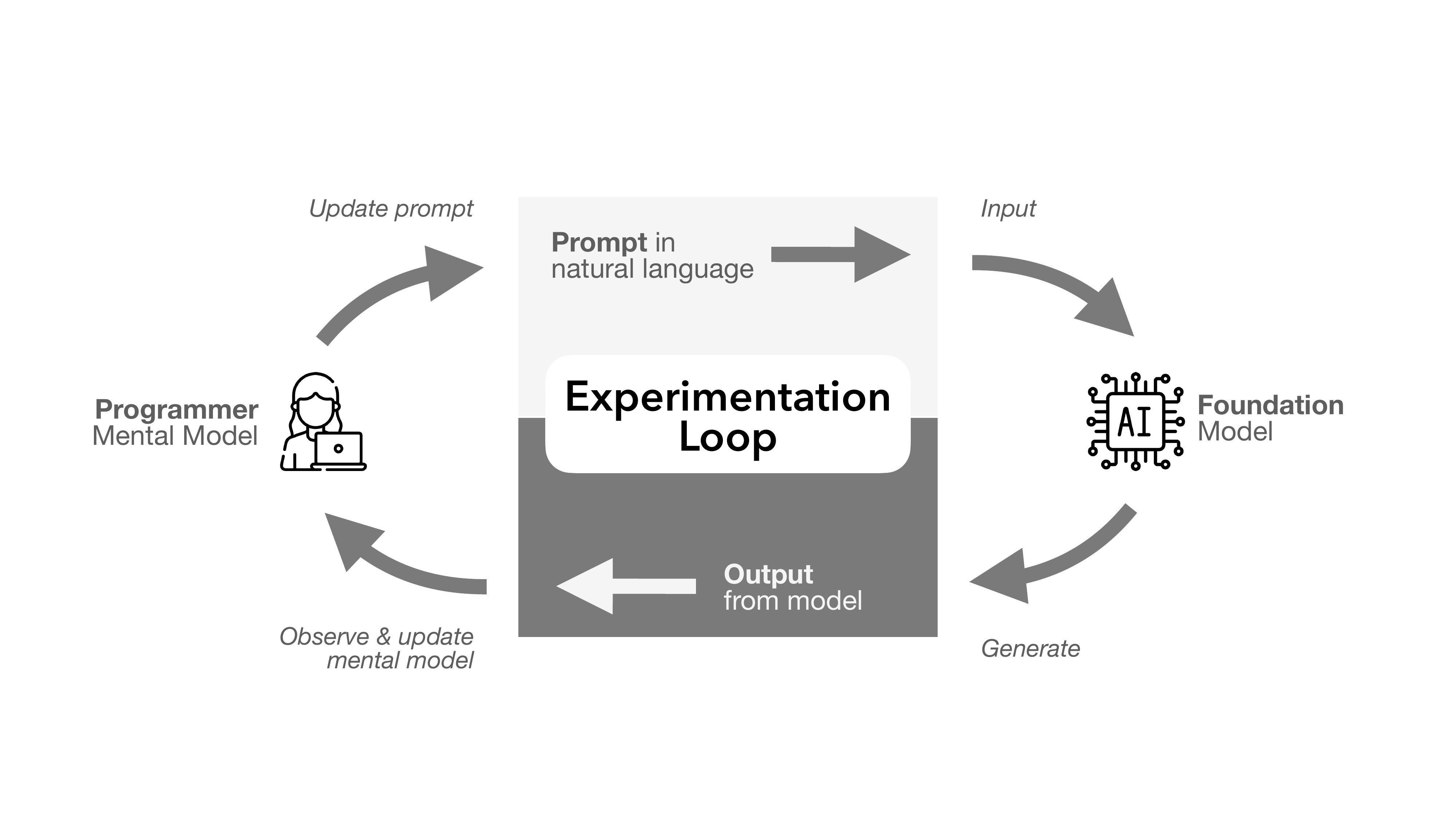} 
\caption{
A programmer uses their prior experience with the foundation model (FM) to construct a mental model of how the FM may perform on the task. 
The programmer uses this mental model to write the prompt.
After observing how the model responds, the developer uses this new information to update their mental model, which influences the next iteration of the prompt.
}
\label{fig:theory}
\end{figure}

\section{Properties of Prompt Programming}
\label{sec:theory}

From the axial coding, we find that prompt programming is the interaction between three entities: the programmer (Section~\ref{sec:programmer}), the foundation model (Section~\ref{sec:model}), and the prompt (Section~\ref{sec:prompt}).
We provide an overview of this process in Figure~\ref{fig:theory}.
First, a programmer uses their prior experience with the FM to construct a mental model of how the FM might perform on the task.
This mental model informs how the prompt is written.
After the prompt is run, the developer observes the generated output to create a hypothesis or new belief about the model's behavior, which then is used to update the developer's mental model.
The updated mental model then influences the next prompt version.
Thus, prompt programs are both very \emph{flexible} in terms of accepting a range of inputs, but also very \emph{overfit} as they are the product of the specific FM being used, the task at hand, and observations of the developer.

\subsection{Programmer}
\label{sec:programmer}
During prompt programming, the programmer spends significant effort developing a mental model of how the FM might behave on the task.

\paragraph{\textbf{Programmers must develop a mental model about the FM's behavior on the prompt to predict how it may behave (\obsGroup{\faLightbulb}{{\#1}}}).}
Technology users form mental representations of interactions with devices, which can be developed by using the technology~\cite{staggers1993mental}.
Research shows that understanding the abilities of a machine learning (ML) model is important but challenging to build ML-enabled systems~\cite{giray2021software, wolf2020sensemaking, dolata2023making, arpteg2018software}, and this applies to FMs as well~\cite{arawjo2023chainforge, jiang2022promptmaker, wu2022promptchainer, hassan2024rethinking}.
Just as developers form mental models of code by exploring the codebase or talking to colleagues~\cite{latoza2006maintaining}, participants described developing a mental model of the FM's behavior on the prompt by running the prompt.
All participants developed their mental model by observing the FM's performance, either by examining individual examples (P6, P7, P8, P9, P10, P12, P14, P15, P16, P17, P18, P19, P20) or looking at metrics (P4, P5, P7, P8, P9, P11, P14, P15, P16, P18, P19).
Based on these observations, the participant formed a hypothesis or belief about the model's behavior and updated the prompt accordingly.
This often included creating a new explicit instruction, \pquote{guideline}{8}, \pquote{rule}{14}, or \pquote{mandate}{20} in the prompt to address the observation, as noted by~\citet{zamfirescu2023herding}.

The programmer's mental model was the accumulated beliefs about the FM's behavior, which could differ or even contradict between individuals.
For example, while some participants (P1, P4, P5, P6, P9, P15, P16, P18, P20) included examples in the prompt for few-shot learning~\cite{brown2020language} to achieve better results, others (P7, P13, P19) believed that the inclusion of examples caused the prompt to overfit to those examples: \pquote{[We did not] use few-shot prompting because...it over-indexes on the examples. After a while, you start seeing very repetitive and monotonous output}{7}.
This overfitting of examples has been found in the NLP literature~\cite{liu2023pre} and has also been observed in practice~\cite{jiang2022promptmaker}.

\paragraph{\textbf{Programmers' mental models are not reliable (\obsGroup{\faLightbulb}{{\#2}}}).}
Mental models are known to be incomplete and inaccurate~\cite{staggers1993mental}.
Due to the lack of transparency of neural models~\cite{arpteg2018software} like FMs in prompt engineering~\cite{mishra2023promptaid}, participants mentioned that they were not always confident in their developed mental models (P2, P4, P5, P6, P8, P9, P10, P15, P17, P18, P19): \pquote{It's a black box model. Nobody knows what's going on inside. It's a science of faith}{17}.
Furthermore, the stochastic nature of FMs~\cite{bender2021dangers} made it difficult for participants to predict the model's behavior, as found in prior work~\cite{jiang2022promptmaker}: \pquote{I have never been confident [in predicting an FM's behavior]...and I don’t think I ever will}{10}.

Some participants (P1, P3, P9, P13, P19) had the FM generate explanations to understand its behavior.
Whereas in NLP literature, these explanations are not \emph{faithful} to the FM (i.e., accurately reflecting the model's actual reasoning process)~\cite{jacovi2020towards}, participants felt that it helped them understand the model's behavior: \pquote{You can reason about [the LLM] if you can see [its] thought process}{19}.
 
\paragraph{\textbf{Programmers use external knowledge sources and prior experience to build their mental model (\obsGroup{\faLightbulb}{{\#3}}}).}
Participants (P1, P2, P4, P11, P12, P15, P16, P18) developed their mental model of the FM via \pquote{prompt intuition}{15} (i.e., prior experience with FMs): \pquote{I knew what works generally just being in the field}{8}.
The literature also points to external knowledge sources as helpful for prompt programming~\cite{parnin2023building, jiang2022promptmaker}.
Participants used official resources, such as guidebooks from organizations or AI companies such as OpenAI (P6, P10, P14, P15), online courses (P7), research papers (P1, P2, P5); informal resources, like blogs, articles (P5, P10, P16, P20), and posts on social networks (P5, P16, P20) from \pquote{everyday users}{20}; expert colleagues (P1, P14, P15); and other example prompts (P16, P18).

Although these resources were helpful, the participants (P2, P10, P14, P15, P16, P20) expressed doubts about whether they improved the prompt: \pquote{I don't know if the best practices are the best. They are there for a reason, but sometimes they just don't actually make that much of a change. There's a reason the [expert] is saying that... I just take it they know better than me}{14}.
Even company-specific resources did not help: \pquote{Even with following best practices...[from] the company, some of them just straight up didn't work}{10}.
Thus, some participants felt that there were no such things as best practices: \pquote{There are no standards for what's a good prompt or what's a bad prompt. It all seems like a haphazard science of just adding a word and maybe it will look better}{17}.

\subsection{Foundation Model}
\label{sec:model}
The FM on which the prompt is run influences how the prompt is written.

\paragraph{\textbf{Each FM has its own set of qualities and capabilities (\obsGroup{\faLightbulb}{{\#4}}}).}
Study participants ran their prompts on models that varied in capabilities (see Table~\ref{tab:participants}).
Some qualities were explicit or obvious.
This included the types of input the model accepted (e.g., images and text (P9)), output generated (e.g., \pquote{GPT-3.5 model's JSON output [mode]}{7}), context window size (P2, P3, P8, P19), prompt formatting requirements (P6), generation speed (P4, P11, P15, P19), and model privacy (P4).

Some capabilities were latent and were discovered by interacting with the model, as noted in previous work~\cite{arawjo2023chainforge, jiang2022promptmaker}.
The literature suggests that FMs have latent capabilities, as FM performance can improve on some tasks and regress on others over time~\cite{chen2023chatgpt, ma2024my}.
Study participants found some differences between FMs to be qualitative: \pquote{[One] model was capturing some instructions in the prompt in a much more focused manner, whereas the [other] was a little more lax}{15}.
In our sample, participants (P1, P4) observed a quality difference in open-source models (e.g., Llama~\cite{touvron2023llama}) compared to closed-source models (e.g., GPT-4~\cite{achiam2023gpt}): \pquote{Getting [the prompt] to work on a smaller model was very frustrating when with a zero-shot prompt, GPT-4 would almost achieve a hundred percent performance}{4}.
Participants (P1, P9) also noted that some FMs were unable to complete the task: \pquote{[The model was] generating garbage, and [after 30 iterations] I concluded the model is trash}{1}.

\subsection{Prompt}
\label{sec:prompt}
Since prompts are the product of the programmer's observations about model behavior and the qualities and capabilities of the FM, prompts can be fragile, and sensitive to small details.

\paragraph{\textbf{Minute details in the prompt matter (\obsGroup{\faDiceSix}{{\#5}}}).}
Participants developed techniques to influence the FM to generate the desired output.
In addition to using techniques from the literature (e.g., assigning personas, selecting a prompting strategy, and providing data context)~\cite{openai2024prompt, lin2024write}, participants used a variety of subtle strategies like using specific phrasing for clarity or generality (P5, P13, P14, P16, P20); repeating phrases for emphasis (P4, P6, P10, P20); avoiding negations (P2, P12); or adding specific characters, such as emojis, to encourage their appearance (P6, P10).
Some participants noted strategies to visually organize the prompt, such as using numbered or bulleted lists (P8, P9, P10, P16) as well as new lines and spacing (P1, P3, P5, P11, P12, P18).
Others used techniques to emphasize specific parts of the prompt, like bolding (P10, P18) and capitalization (P7, P10, P12, P20).

The participants (P10, P12) also decided on creative details to include in the prompt for better performance.
For example, to prevent the FM from leaking prompts, one participant \pquote{bullied GPT}{10}: 
\pquote{We have a very light threat in there saying, `If you share any of this information...you [and] I will get in a lot of trouble and it will cause great career harm.'}{10}.
Another participant summarizing code told the FM to explain the code \pquote{`so a sixth grader who doesn't understand programming can understand it'. The explanations...were verbose, so I added, `sixth graders don't really like reading' and for some reason, that got it to be shorter}{12}.

Finally, participants noted that the input and output data format impacted the prompt performance.
Consistent with previous studies~\cite{parnin2023building, liu2024we}, participants selected structured and standardized data formats, such as JSON (P3, P7, P9, P10, P11, P18), Markdown (P2, P7, P12) and XML tags (P1, P3, P7, P10, P20).
The participants would change the data representation to achieve better performance (P4, P13, P20): \pquote{[What worked better is] describing the APIs as though they are Python functions and having the documentation in a similar format to publicly available Python functions}{13}.

\paragraph{\textbf{Prompts are finicky and fragile (\obsGroup{\faDiceSix}{{\#6}}}).}
As noted in the NLP literature~\cite{jiang2020can, ye2024prompt} and empirical studies on prompting~\cite{liu2023pre, zamfirescu2023herding}, the performance of a prompt varies based on its language.
In practice, participants found prompts to be \pquote{finicky}{4}, \pquote{fragile}{20}, \pquote{sensitive}{1, P13}, and \pquote{temperamental}{15}, since prompt performance could suddenly worsen during development, as corroborated in previous studies on prompt engineering~\cite{parnin2023building, zamfirescu2023herding, zamfirescu2023conversation, jiang2022promptmaker, wu2022promptchainer, hassan2024rethinking}.
This could be due to a FM's ability to follow instructions, which can change over time~\cite{chen2023chatgpt}.
For instance, participants noted performance regressions (P2, P8, P10, P12, P15, P16, P19, P20), often caused by adding in \pquote{a bunch of new requirements}{2} that would cause previous instructions to suddenly \pquote{deactivate}{20} or be \pquote{disregarded}{8} in the new version: 
\pquote{If I provide multiple guidelines, [the FM] would follow some and disregard others, especially when they're conflicting or overlapping in nature}{8}.

While previous NLP studies point to prompts being transferable between FMs~\cite{liu2023pre}, participants found prompts to be \pquote{finicky between models}{4}, causing the prompt to break and require additional development effort (P4, P10, P15, P18): \pquote{At one point, we had to do...a minor model version change... That broke almost everything. We had to rewrite so many prompts}{10}.
This is because a FM change could result in \pquote{a different format... I have a regex converting [the output]...and the whole thing gets broken}{18}.
This has also been observed in existing prompt engineering studies~\cite{dolata2024development}.

\begin{table*}
  \centering \footnotesize
\caption{A comparison between traditional software development and prompt programming in terms of \added{prompt programming} activities.}
\label{tab:comparison}
\begin{tabular}{p{0.16\linewidth}|p{0.37\linewidth}|p{0.37\linewidth}}
\toprule
& \textbf{Traditional software development} & \textbf{Prompt programming} \\
\midrule
\multirow{4}{0.16\linewidth}{\textbf{Requirements} \\ \added{(Section~\ref{sec:requirements})}} & • Expressed outside of code & \hangindent=1em • Expressed in a prompt, which is also the implementation \added{(\obsGroup{\faFileCode}{{\#7}})} \\
& & \hangindent=1em • Dependent on the performance of the FM on the task \added{(\obsGroup{\faLightbulb}{{\#8}})} \\
\midrule
\multirow{2}{0.16\linewidth}{\textbf{Design} \\ \added{(Section~\ref{sec:design})}} &  • Base component: Class/function &  • Base component: Single prompt \added{(\obsGroup{\faFileCode}{{\#9}})} \\
&  &  • Chained prompts are tightly coupled \added{(\obsGroup{\faLightbulb}{{\#10}})} \\
\midrule
\multirow{4}{0.16\linewidth}{\textbf{Implementation} \\ \added{(Section~\ref{sec:implementation})}} & \hangindent=1em • Write a program that implements a specification in code & \hangindent=1em • Write a prompt that is a natural language specification that when \added{run} on a \added{FM}, implements the specification \\
& & • Focus on rapid experimentation \added{(\obsGroup{\faLightbulb}{{\#11}})} \\
\midrule
\multirow{3}{0.16\linewidth}{\textbf{Debugging} \\ \added{(Section~\ref{sec:debugging})}} &  • Fault can be isolated to a single line of code &  • Fault localization is not certain \added{(\obsGroup{\faDiceSix}{{\#12}})} \\
 &  \hangindent=1em • Can use debugging tools for more systematic debugging &  \hangindent=1em • No debugging tools; reliance on shotgun debugging \added{(\obsGroup{\faDiceSix}{{\#12}})} \\
\midrule
\multirow{2}{0.16\linewidth}{\added{\textbf{\mbox{Data Curation}} \\ (Section~\ref{sec:data-curation})}} & • \added{N/A} & • Focus on gathering \added{a} representative \added{dataset that reflects the task data} \added{(\obsGroup{\faClipboardCheck}{{\#13}})} \\
\midrule
\multirow{3}{0.16\linewidth}{\added{\textbf{Evaluation} \\ \added{(Section~\ref{sec:evaluation})}}}& • Focus on code coverage metrics & • Focus on \added{qualitative evaluation} \added{(\obsGroup{\faClipboardCheck}{{\#14}})} \\
& \hangindent=1em • Testing at varying scopes (e.g., unit testing vs. integration testing) & • \hangindent=1em Testing at varying scopes (e.g., testing single prompt vs. testing chained prompts) \added{(\obsGroup{\faClipboardCheck}{{\#15}})} \\
\bottomrule
\end{tabular}
\end{table*}

\section{Prompt Programming Process}
\label{sec:process}
The prompt programming process has significant differences compared to standard software development (see Table~\ref{tab:comparison}).
We identified six activities in prompt programming from axial coding:
requirements (Section~\ref{sec:requirements}), design (Section~\ref{sec:design}), implementation (Section~\ref{sec:implementation}), debugging (Section~\ref{sec:debugging}, data curation (Section~\ref{sec:data-curation}), and evaluation (Section~\ref{sec:evaluation}).

\subsection{Requirements}
\label{sec:requirements}
In traditional software, requirements are separate from code, while in prompt programming, they are stated in the prompt and may evolve with the behavior of the FM.
As in prior work~\cite{furmakiewicz2024design, parnin2023building}, participants noted functional and non-functional requirements, such as usability (P13, P15), reliability (P7, P19), latency (P4, P11, P15, P19), safety (P7, P10, P13, P19), privacy (P4, P13, P14), availability (P4), and cost (P5, P8, P10, P11, P14, P16, P19, P20).
Companies found safety and privacy important, as noted in the literature~\cite{hassan2024rethinking, patel2024state}.

\paragraph{\textbf{Assumptions of the requirements must be explicitly stated (\obsGroup{\faFileCode}{{\#7}}}).}
At times, the FM performed an incorrect action \textit{not} because of an FM error, but due to an under-specified or conflicting requirement in the prompt (P3, P4, P5, P6, P8, P12, P19).
Thus, prompt programming required explicitly stating assumptions: \pquote{When you're interacting with a person, you make all these assumptions, but everybody can read between the lines... The more you design prompts, the more you realize that it's not the same as communicating with another person. The model needs more specific instructions}{3}.
However, predicting which assumptions to elaborate on was challenging, as \pquote{it's hard to tell when the model could go wrong}{3}: \pquote{I think observation is like 80\% [of the requirements]. I knew what I wanted 20\% of the time}{8}. 
Thus, \pquote{capturing all the rules...is very difficult in the start, and just generalizing over all the use cases can be very difficult}{19}.

\paragraph{\textbf{Requirements can evolve as the capabilities of the FM are discovered (\obsGroup{\faLightbulb}{{\#8}})}.}
Similar to building ML-enabled systems~\cite{wan2019does}, prompt requirements could change as the FM's capability on the task was discovered.
This included making new trade-offs between non-functional requirements, such as safety vs. quality~\cite{furmakiewicz2024design} (P7): \pquote{Our requirements did change as to how the product was perceived... So, striving for that balance between the response quality and the latency induced quite a few iterations}{15}.
The participants also defined new features after observing how the FM performed on the task (P2, P4, P13, P18, P20): \pquote{I do add in requirements as I iterate. `Oh, I didn't know that you would be able to do this. Let's see how much I could push this functionality'}{2}.

\subsection{Design}
\label{sec:design}
Just as traditional software could be decomposed to base components of classes and functions, complex prompts could be decomposed into multiple tightly coupled prompts.

\paragraph{\textbf{Developers make decisions about how prompt programs should be composed and decomposed (\obsGroup{\faFileCode}{{\#9}}).}}
Participants decomposed prompt programs into smaller components.
For a single prompt, the participants decomposed the prompt into subcomponents such as subsections (P1, P2, P5, P6, P7, P8, P13, P20).
One participant noted \pquote{it is important to structure the prompt in a way that makes sense to you. Otherwise, you're not able to maintain it going forward}{13}.

Prompts can be composed or ``chained'' together for greater capabilities~\cite{liu2023pre, wu2022promptchainer, wu2022ai, arawjo2023chainforge}.
The participants decided how to decompose the prompts: \pquote{We know what the complex task looks like, and then our job is to break that down}{9}.
Each prompt had its own responsibility: \pquote{If you are creating a robot, the motion of the hand in any scenario could be a responsibility [and] the movement of the legs could be another}{19}.
Similarly to pure code systems, prompts acted as a single module with its own concern~\cite{dijkstra1982role} that could be composed into a more complex system.

\paragraph{\textbf{Dependent prompts are tightly coupled (\obsGroup{\faLightbulb}{{\#10}})}.}
As observed in other studies~\cite{wu2022promptchainer, dolata2024development}, dependent prompts posed challenges since developers needed to form mental models of how the prompts performed together, rather than individually: \pquote{Two changes that do really well separately could merge and become a terrible experience for unexplainable reasons}{13}.
This is similar to engineering ML-enabled systems, whose components have tight coupling~\cite{wan2019does}.
Participants (P8, P12, P13, P19) found that integration resulted in breakage and required significant prompt changes to work with each other.
When working individually, P19 performed regular integration testing of multiple prompts: \pquote{[Integration testing is] highly correlated with how it's going to perform in a production environment. In the end, we are going to integrate everything}{19}.
However, collaboration could pose challenges, since programmers may be unable to test with other in-progress prompts.
P13 reported changing how she would \pquote{split up the work}{13} with her colleagues and felt a \pquote{rush to get my PR [(pull request)] in before the other one does}{13} to avoid the additional work associated with integration.
Participants tried to address this challenge by allowing APIs like LangChain (P4, P9, P10) or LangGraph (P19) to handle orchestration, as noted in previous work~\cite{wu2022promptchainer, dolata2024development}.

\subsection{Implementation}
\label{sec:implementation}
In prompt programming, natural language prompts serve as both the specification and implementation, while traditional software implements specifications in code.
As mentioned in Section~\ref{sec:theory}, prompt programming involves constant \pquote{experimentation [compared to] traditional software development}{15} and is a process of \pquote{trial and error}{3, P7, P10, P15}.

\paragraph{\textbf{Prompt programming is rapid and unsystematic \obsGroup{\faLightbulb}{{\#11}}}).}
To test the prompt's feasibility, participants (P1, P4, P6, P7, P8, P12, P13, P14, P15, P19, P20) started with a basic prompt for the task which included a description of the task and how a human would intuitively perform the task.
Each update could result in unwanted behavior and require prompt updates, thus restarting the update-test loop.
While developing a mental model of the FM (Section~\ref{sec:programmer}), the prompt became more complex as the programmer refined it and observed the FM's behavior.
This process was \pquote{unscientific}{4}: \pquote{I would try a lot of different things on a prompt and it's not a very well thought-out procedure}{6}. 
It also occurred even with existing prompt programs (P14).
Previous work has characterized prompt engineering as highly iterative~\cite{parnin2023building, zamfirescu2023conversation, zamfirescu2023herding, arawjo2023chainforge, zamfirescu2023johnny, dolata2024development, lin2024write, liu2024we}, similar to the experimental nature of engineering ML-enabled systems~\cite{giray2021software, nahar2022collaboration, nahar2023meta, wan2019does, wolf2020sensemaking}.

During experimentation, \pquote{the speed and the velocity of making changes [was] high}{19}.
The literature indicates that finding an optimal prompt is challenging~\cite{mishra2023promptaid, arawjo2023chainforge}, as the rapid experimentation produces many outputs that are difficult to make sense of~\cite{jiang2022promptmaker, jiang2023graphologue, gero2022sensemaking, liu2024we}.
Participants created many prompt versions to test new approaches (P1, P5, P19), add functionality (P2, P6, P13, P16), fix bugs (all participants), experiment with new models or hyperparameters (P4), or incorporate feedback after a code review (P1, P14, P16).
Prompt versions often branched off from each other (P4, P6, P12, P13, P15, P17, P19, P20), but could be reverted (P4, P5, P8, P10, P15, P16, P20).

As noted in earlier studies~\cite{zamfirescu2023herding}, changes were less drastic as the modifications had less of an effect on the prompt, leading \pquote{prompt engineering [to feel] like a waste of time}{5}: \pquote{If we graphed the number of...changes over the number of iterations, the gradient would be pretty high during the first iterations. Then it flattens quite a bit...because the prompt has matured...such that we don't need...big changes}{15}.
P16 avoided testing small changes and preferred to batch them to reduce costs: \pquote{For small changes, we will wait and merge this evaluation into the next time we conduct it}{16}.

Consistent with prior studies in prompt engineering~\cite{parnin2023building, jiang2022promptmaker} and building ML-enabled systems~\cite{nahar2022collaboration, nahar2023meta, arpteg2018software, giray2021software}, the AI's nondeterminism makes it \pquote{hard to measure overall progress}{6}: \pquote{You can think you have something that's really good, and then try it out on a completely different set of inputs and it's just terrible. A lot of this is feeling of making progress and then backtracking}{10}.
Participants developed strategies to address this, such as making small changes (P2, P4, P8, P15) or viewing the evaluation metrics computed on a large number of examples (P4, P18): \pquote{I feel like I wasn't that confident, other than the fact that our intrinsic numbers went up}{4}.

The study participants used a variety of tools.
GUI-based tools (P2, P4, P5, P6, P8) such as GPT playground and notebooks (P1, P3, P5, P18, P19) enabled rapid prototyping; text editors and IDEs (P1, P11, P14, P17, P20) revealed formatting, like newline characters; and FMs helped inspire or generate parts of the prompt (P2, P18).
The prompts were manually transferred between environments, usually by copy-pasting (P1, P2, P3, P4, P5, P8).
This introduced issues like formatting errors (P1, P5) and inconsistent behavior between using the GUI versus the API environments (P5).

Due to rapid experimentation, the participants (P3, P18, P19) struggled to recall previous versions and their output.
Some tried to address this by tracking versions of the prompt as found in~\citet{parnin2023building}'s study by using Git (P5, P6, P7, P10, P16, P17, P18) or even custom systems.
These systems varied in sophistication, from a text document tracking all prompts and outputs (P1, P2, P5, P12) and arbitrary file-naming systems (P9, P12, P19), to spreadsheet tables (P1, P4, P5) and custom-built tools (P8).
However, participants found it difficult to retrieve previous prompt versions (P4, P12) and confused versions due to subtle differences between versions (P9, P12).

\subsection{Debugging}
\label{sec:debugging}
Participants observed defects like hallucinations (P9, P14, P15, P17, P19) or not following directions (P2, P3, P4, P6, P8, P9, P10, P12, P18, P19, P20).
Unlike traditional software, where debugging tools can isolate faults to a line of code, prompt debugging is unsystematic due to FM stochasticity.

\paragraph{\textbf{Fault localization is difficult (\obsGroup{\faDiceSix}{{\#12}}}).}
To fix errors, the participants debugged prompts, which is known to be challenging~\cite{jiang2022promptmaker, dolata2024development}.
The stochastic and opaque nature of FMs made it impossible to determine the source of the defect: \pquote{Coming from a software engineering background, you...want to set breakpoints...looking at the results step by step. There's no such mechanism for prompts}{13}.
One challenge was to know how to change the prompt to fix the defect~\cite{mishra2023promptaid}: \pquote{I don't even know what to change in my prompt to get there}{5}.
Participants engaged in shotgun debugging by trying random changes to fix the defect (P2, P7, P8, P10, P15, P18): \pquote{In most cases, it is like hit and trial. So maybe I will try a different prompt...[or] example}{18}.
Thus, prior experience and intuition were helpful in debugging (P2, P4, P15, P16).
Study participants also described strategies to gain more confidence in understanding what could fix the error.
This included having the FM generate reasoning (P1, P3, P9, P13, P19); testing small changes (P12); and restricting instructions to specific parts of the prompt (P14, P20).
However, the change did not always fix the error (P3, P4, P5, P6, P15, P16, P18, P20), as it could introduce new errors (P3, P4, P7), as documented in previous work~\cite{parnin2023building}.

\subsection{Data Curation}
\label{sec:data-curation}
To develop prompt programs, programmers must create  datasets, a process referred to as ``data curation''~\cite{qian2024understanding}.
This became a central activity that is not present in traditional software engineering.

\paragraph{\textbf{Programmers need to find representative data for the task (\obsGroup{\faClipboardCheck}{{\#13}}}).}
Using FM-based components shifts the focus to finding high-quality data, consistent with previous work~\cite{hassan2024rethinking, nahar2022collaboration, arpteg2018software, dolata2023making, giray2021software, wan2019does, nahar2023meta}.
To evaluate the prompt and provide examples for few-shot learning~\cite{brown2020language}, data curation was paramount.
Since many participant tasks were custom or specific, the existing benchmarks were not sufficient except for more common tasks like jailbreaking (P19).
Study participants created their own datasets by mining data from the internet or from their organization, as well as annotating data (P1, P3, P4, P5, P6, P7, P8, P11, P13, P14, P15, P17, P18, P20).
The challenge was to find representative data that worked in practice, as prompt programmers needed \pquote{to predict how people are going to use [the application]}{20}.
Datasets did not always work in practice due to \emph{training-serving skew} (i.e., when training data does not generalize to production data)~\cite{nahar2022collaboration, nahar2023meta, wan2019does}:
\pquote{You can test with your own dataset, but at the end of the day, you’re still not completely sure how good the output is}{7}.

The participants ensured diverse inputs by categorizing the inputs and selecting examples for each category (P1, P8, P9, P15, P19, P20). 
Previous work indicates that the nondeterminism of ML models poses difficulties in evaluation~\cite{wan2019does, hassan2024rethinking}. 
The participants addressed this by paraphrasing the inputs (P16, P18, P19, P20).
Despite their efforts, the datasets did not capture the input distribution, so the participants relied on several methods to obtain more data like using FMs to generate examples (P12, P18, P19); asking colleagues to stress test, \pquote{data bash}{13}, or red team~\cite{furmakiewicz2024design} the prompt (P7, P10, P13, P17, P19); and obtaining user feedback (P7, P10, P13, P15, P17, P19).

As found in other work~\cite{arpteg2018software}, participants struggled to create datasets.
Despite the emphasis on safety and fairness in the literature~\cite{giray2021software, arpteg2018software}, prompt programs performed worse on culturally diverse audiences during deployment (P7, P10, P13).
This reflects NLP research, where FMs and datasets align with certain demographics (i.e., Western, college-educated, and younger populations) more than others~\cite{santy2023nlpositionality, pei2023annotator, joshi2020state}.
Data labeling was also challenging, as annotations could be of lower quality: \pquote{The [annotations from some colleagues] would be incorrect...they have some limitations on how much they know, [so] they don't know what the expected outcome is}{13}.
Meanwhile, the correct label could be unclear (P3, P13): \pquote{When do you classify as [Label A]? It was fuzzy on how you would interpret [the labels]}{3}.
This reflects the challenges of handling annotator disagreement and subjectivity in NLP research~\cite{aroyo2015truth}, where ground truth labels can be subjective.

\subsection{Evaluation}
\label{sec:evaluation}
Both traditional software and prompt programs involve testing at multiple scopes.
Traditional software uses code coverage, while prompt programming often relies on qualitative evaluation.

\paragraph{\textbf{Evaluating prompt programs requires assessing qualitative constructs (\obsGroup{\faClipboardCheck}{{\#14}}}).}
The participants had custom tasks that did not have a quantified notion of correctness, making the evaluation difficult~\cite{parnin2023building, dolata2024development}.
For building ML-enabled systems, many evaluations are quantitative metrics~\cite{wan2019does, giray2021software, nahar2022collaboration}, while prompt engineering uses manual inspection~\cite{gero2022sensemaking, mishra2023promptaid}.
Participants performed manual tests by qualitatively assessing FM outputs (P2, P4, P6, P7, P8, P12, P13, P14, P16, P17, P18, P19, P20) as it was difficult to \pquote{quantify the concept of good}{10}.
For example, P2, who created a voice memo summarization tool, did the following to test his application: \pquote{I would...talk to my phone for like 10 minutes just to fine-tune and see, `Does the summary make sense?'}{2}.

Manual testing was difficult to scale and quantify progress (P6, P15): \pquote{It is extremely hard to realize the impact of the [changes] without...a person who is very finely checking out the differences}{15}, especially in the face of regressions.
Participants supplemented this with quantitative metrics via scaled programmatic testing (P1, P5, P8, P9, P11, P13, P14, P16, P18): \pquote{I run [my prompt] manually on 3 to 4 examples... [When] I have a proper evaluation setup, I run on 50 to 100 cases. Those 50 cases make me more confident}{18}.
This involved using other FMs to generate scores for qualitative constructs (P4, P7, P9, P10, P15, P19), like safety~\cite{furmakiewicz2024design}: \pquote{I have a scoring mechanism with...an evaluation agent [where the] score is from 0 to 100}{19}.
However, FM-based evaluations could also introduce errors: \pquote{[The evaluation agent] itself is an LLM...So 10\% of the time, if the test case is failing, it's because the evaluation agent wasn't able to score it correctly}{19}.

\paragraph{\textbf{Testing occurs at different scopes (\obsGroup{\faClipboardCheck}{{\#15}}}).}
Participants tested their prompts at varying scopes.
For applications with multiple prompts, participants tested individual components of the prompt program (P8, P16, P19, P20), akin to unit testing: \pquote{Usually, each test case that we had written was aimed at testing one specific part of the prompt}{20}.
The participants also described performing integration tests (P2, P8, P9, P11, P12, P13, P14, P16, P18, P19): \pquote{We evaluated each of [the prompts] separately, and now we are evaluating that as a collection}{11}.
However, this type of testing was difficult, as combining prompts could yield unexpected interactions.

\section{Discussion}
We discuss how our findings relate to prompt engineering and traditional software engineering (Section~\ref{sec:comparison}), their implications (Section~\ref{sec:implications}), and takeaways for various stakeholders (Section~\ref{sec:recommendations}).

\subsection{Prompt Programming: Where Software Engineering Meets Prompt Engineering}
\label{sec:comparison}
Prompt programming combines elements of traditional software engineering and prompt engineering.
Like traditional programs, prompt programs must handle arbitrary inputs, often requiring the decomposition of complex tasks into modular components~\cite{liu2023pre, wu2022promptchainer, wu2022ai, arawjo2023chainforge} (\obsGroup{\faFileCode}{{\#9}}) that require different forms of testing (\obsGroup{\faClipboardCheck}{{\#15}}). 
At the same time, like prompt engineering~\cite{strobelt2022interactive, zamfirescu2023conversation, zamfirescu2023herding, arawjo2023chainforge, zamfirescu2023johnny, dolata2024development, lin2024write, liu2024we}, interactions with opaque FMs drive rapid and unsystematic experimentation (\obsGroup{\faLightbulb}{{\#11}}), a practice less common in traditional software engineering.
However, some challenges are particularly prominent in prompt programming, which we discuss below.

\paragraph{Emphasis on trial-and-error (\obsGroup{\faLightbulb}{{\#11}})} 
Prompt programmers must develop a mental model of FM behavior on diverse inputs~\cite{arawjo2023chainforge, jiang2022promptmaker, wu2022promptchainer, hassan2024rethinking} by running the prompt (\obsGroup{\faLightbulb}{{\#1-\#4, \#8}}). 
This reliance on execution stems from the inherent opaqueness of FMs~\cite{arpteg2018software, mishra2023promptaid}, which lack the deterministic and interpretable properties of most conventional software. 
Although understanding program behavior in traditional software engineering can be challenging~\cite{maalej2014comprehension}, this issue is especially critical in prompt programming, where reading code or consulting documentation alone is insufficient.

\paragraph{Quality assurance via manual testing and dataset curation (\obsGroup{\faClipboardCheck}{{\#13-\#14}})} 
Handling arbitrary inputs in prompt programs also requires extensive testing~\cite{strobelt2022interactive} (\obsGroup{\faClipboardCheck}{{\#13-\#15}}) and a clear articulation of the implicit and explicit requirements (\obsGroup{\faFileCode}{{\#7}}) to ensure reliability and generalizability. 
Consequently, prompt programming places greater emphasis on dataset curation (\obsGroup{\faClipboardCheck}{{\#13}}) and subjective evaluation~\cite{gero2022sensemaking, mishra2023promptaid} (\obsGroup{\faClipboardCheck}{{\#14}}), both less prevalent in traditional software engineering and prompt engineering.

\paragraph{Handling nondeterministic program behavior (\obsGroup{\faDiceSix}{{\#5-\#6, \#12}})}
Arbitrary inputs make the stochastic nature of FMs pronounced in prompt programming~\cite{jiang2020can, ye2024prompt, liu2023pre, zamfirescu2023herding} from implementation (\obsGroup{\faDiceSix}{{\#5-\#6}}) to debugging (\obsGroup{\faDiceSix}{{\#12}}). 
While nondeterministic behavior in traditional software (e.g., flaky tests) can be identifiable and mitigable (e.g., test order dependencies)~\cite{luo2014empirical}, it is harder to diagnose in prompt programming given the lack of understanding of FM behavior.

\paragraph{Implications for evolving FM capabilities} 
With the rapid evolution of FM capabilities, prompt fragility (\obsGroup{\faDiceSix}{{\#6}}) may become less relevant over time. 
Findings related to stochasticity will likely persist, albeit to a lesser extent, since natural language ambiguity will still make prompt details influential (\obsGroup{\faDiceSix}{{\#5}}) and contribute to uncertainties in fault localization (\obsGroup{\faDiceSix}{{\#12}}), even as models improve.
Regardless of FM performance, we expect programmers to continue to form mental models of FM behavior (\obsGroup{\faLightbulb}{{}}), test prompt program behavior (\obsGroup{\faClipboardCheck}{{}}), and program in natural language (\obsGroup{\faFileCode}{{}}).

\subsection{Implications}
\label{sec:implications}

\paragraph{Prompt programmers are both ML practitioners and software developers}
Developing ML-enabled systems requires a variety of expertise such as software engineering, data science, and math~\cite{wan2019does, giray2021software}.
This expertise is captured across team members, which can introduce collaboration challenges~\cite{nahar2022collaboration}.
Though prompt programming involves both ML and programming knowledge, this expertise is now centralized in a single individual.
This could reduce friction between different roles.
However, in addition to traditional software engineering expertise like debugging, testing, or knowledge in specific programming languages~\cite{liang2022understanding, baltes2018towards}, prompt programmers must develop new skills as they engage in novel activities like building datasets and debugging model errors with data~\cite{qian2024understanding}.

Since prompt programmers are essentially ML practitioners, these programmers encounter many of the long-standing challenges contended by the ML and NLP communities, like model drift~\cite{chen2023chatgpt}, language and cultural biases in datasets and models~\cite{joshi2020state, santy2023nlpositionality, pei2023annotator}, and annotator subjectivity~\cite{aroyo2015truth}.
Close collaboration between the software engineering and ML communities is paramount to inform techniques that address these challenges in prompt programming.

\paragraph{Prompt programmers are inundated with information}
The rapid iteration involved in prompt programming means that prompt programmers must make sense of a deluge of information.
For each iteration, the participant must track the prompt, the change being made, the FM's output, and associated metrics.
However, participants reported having difficulty making sense of all of this information (P3, P6, P18, P19).
This parallels exploratory programming, which is also highly iterative and unsystematic in nature~\cite{kery2017exploring} and requires tracking of many artifacts, including source code, input data, output data, code snippets, and analysis progress~\cite{kery2017variolite}.

\paragraph{Are the ``best practices'' best practices?}
In our study, participants' mental models on FMs conflicted, such as whether few-shot learning~\cite{brown2020language} was beneficial for prompt performance.
Although the inclusion of examples is listed within OpenAI's official prompting guide~\cite{openai2024prompt}, some of our participants (P7, P13, P19) were skeptical of this practice after observing the FM.
This underscores the importance of the observations of developers in prompt programming, especially in the face of evolving model behaviors~\cite{chen2023chatgpt, ma2024my}.
However, such knowledge and observations are rarely shared, are siloed in informal social media posts~\cite{parnin2023building}, and are not systematically collected and centralized.

\subsection{Takeaways}
\label{sec:recommendations}

\begin{figure}
\begin{tcolorbox}[colback=PAlightblue,colframe=PAblue,title=\textbf{\hspace{1.65em} Prompt Programming Recommendations},left=-12pt,right=8pt,top=4pt,bottom=4pt, fontupper=\small, fonttitle=\small]
      \begin{itemize}
        \item Get hands-on experience with prompt programming to develop a mental model of FM behavior.
        \item Treat prompt programming best practices as a starting point and vet them in the specific context.
        \item Learn data science skills.
        \item Curate diverse and evolving datasets with varying examples that capture the domain.
        \item Record prompt changes and FM behavior in a structured format during each iteration.
        \item Coordinate with colleagues and conduct frequent integration tests for chained prompts.
    \end{itemize}
    \end{tcolorbox}
    \vspace{-0.5\baselineskip}
    \caption{Recommendations for prompt programming for educators and practitioners.}
    \label{fig:recommendations}
\end{figure}

\paragraph{For practitioners and educators}
The lack of standardized best practices and unreliable mental models of prompt programmers pose challenges in teaching and learning prompt programming.  
Future work should investigate methods to effectively teach prompt programming. 
In the meantime, educators should consider the following recommendations, which are summarized in Figure~\ref{fig:recommendations}.

Having prior experience with FMs and \pquote{prompt intuition}{15} made it easier to build prompt programs (\obsGroup{\faLightbulb}{{\#3}}).  
Since this knowledge is tacit, practitioners should focus on obtaining hands-on experience with prompt development to build a preliminary understanding of FM behavior.
Since prompt programming best practices (e.g., from OpenAI) do not work in all situations (\obsGroup{\faLightbulb}{{\#3}}), they should be treated as a starting point.
As our participants showed, these guidelines should be vetted, such as through online communities, and validated within the programmer’s specific context.

Since prompt programming also requires an understanding of data for rigorous testing of prompts (\obsGroup{\faClipboardCheck}{{\#13}}), practitioners should learn data science skills like dataset curation, data cleaning, data shaping, and debugging model errors with data~\cite{kim2016emerging, qian2024understanding}.
Programmers should develop benchmarks with representative data to catch performance regressions and evolve to address \emph{training-serving skew}~\cite{nahar2022collaboration}.
Study participants did so by collecting different categories of data, such as input paraphrasing and linguistically diverse inputs.

Lastly, practitioners should consider changes to their development processes.
Since prompt programming is unsystematic (\obsGroup{\faLightbulb}{{\#11}}), programmers should record iterations in a structured manner, summarizing changes to the prompt and FM behavior to increase systematicness.
Study participants did so by recording different versions and their design rationales in spreadsheets.
Practitioners should also prioritize collaboration due to the tight coupling of chained prompts (\obsGroup{\faLightbulb}{{\#10}}) by performing regular check-ins with colleagues and frequent integration testing. 

\paragraph{For researchers and tool makers}
\pquote{[Prompt programming] will require any...developer to adjust their mindset... 
It is much more scientific and experimental than traditional software development}{15}.
Participants struggled at each step, from requirements to evaluation, as existing developer tools were not designed to handle the challenges of prompt programming.
We follow \citet{hassan2024rethinking} and \citet{parnin2023building} in underscoring the importance of developer tools for prompt programming. 
Such explorations have begun in earnest~\cite[e.g.,][]{mishra2023promptaid, wu2022promptchainer, jiang2022promptmaker, arawjo2023chainforge}.
Future tool ecosystems should support the full range of activities in prompt programming, from interfaces for rapid prototyping, dataset curation, and data sensemaking, to regression testing, prompt chaining, collaboration, and more.
We elaborate on potential prompt programming tool directions below.

Given the large volume of information prompt programmers process per iteration (e.g., examples, prompt modifications, and FM output changes), future work should explore improved methods for versioning and navigating prompt iterations and artifacts.
This could provide rich information on prompt provenance and could be used for automatic prompt program repair.
A key challenge is deciding what to version and how to represent both prompt and FM output changes. 
Because prompt changes are made with varying granularity, from editing single words to changing prompting strategies, research is needed to understand how to visualize these changes between versions.
Meanwhile, research should study new visualizations for FM output changes.
This is challenging as prompt programmers qualitatively compare the same prompt across multiple examples~\cite{gero2022sensemaking}.

Future work is also needed for the reuse of prompts, since the fragility of prompts and the unique qualities of each FM make it difficult to reuse prompts without modification.
Reuse at higher levels of abstraction (e.g., template-based approaches~\cite{white2023prompt, white2024chatgpt}), and community-based solutions for prompt sharing (e.g., ShareGPT~\cite{sharegpt2024sharegpt}, LangSmith Hub~\cite{langchain2024langsmith}, Wordflow~\cite{wang2024wordflow}) could be promising avenues for prompt reuse.
Research is needed to understand how to retrieve relevant prompts.
Prompt programmers could query prompts based on different information, ranging from specific keywords in the input and FM output to general FM behavior or performance metrics.
Another challenge is ensuring that the recommendations can generalize across diverse, situation-specific contexts of different prompt programmers (e.g., identifying optimal wording for specific applications). 

\section{Conclusion}
Foundation models (FMs) enable programmers to write natural language prompts that power AI-driven software features.
We study this phenomenon---prompt programming---via 20 interviews with developers across various contexts, using Straussian grounded theory to analyze the process.
We identified 15 observations about prompt programming barriers.
We find that prompts are the product of the developer's understanding of FM and the developer's observations of the FM's behavior on the task.
We also find strong parallels between the prompt programming process and the traditional software engineering process, but also significant differences.
Our findings inform both researchers developing prompt programming tools and practitioners who use them.

\section*{Data Availability}
To facilitate replication, our supplemental materials are available on FigShare~\cite{supplemental-materials},including our survey instrument, interview protocol, codebook, and the 33 reviewed papers. 

\begin{acks}
We thank our participants for their insights.
We express gratitude to Chenyang Yang, Nadia Nahar, Travis Breaux, and Christian Kaestner for their feedback.
Last but not least, we give a special thanks to Mei \meiicon, an outstanding canine software engineering researcher, for providing support and motivation.
Jenny T. Liang was supported by the National Science Foundation under grants DGE1745016 and DGE2140739.
Any opinions, findings, conclusions, or recommendations expressed in this material are those of the authors and do not necessarily reflect the views of the sponsors.
\end{acks}

\bibliographystyle{ACM-Reference-Format}
\bibliography{acmart}

\end{document}